\documentclass[cits]{PoS}

\newcommand{\mics}{$\mu$m}
\newcommand{\lfir}{$L_{60}$}
\newcommand{\lagn}{$L_{AGN}$}
\newcommand{\ergs}{erg s$^{-1}$}
\newcommand{\smass}{M$_{\large *}$}

\title{Star-formation in active galaxies to z$\sim$2: a perspective from Herschel studies}

\ShortTitle{Star-formation in active galaxies to z$\sim$2: a perspective from Herschel studies}

\author{\speaker{David Rosario}%
        \thanks{Herschel is an ESA space observatory with science instruments provided by European-led Principal Investigator consortia and with important participation from NASA.}\\
       Max-Planck-Institut f\"ur extraterrestrische Physik, Germany\\
       E-mail: \email{rosario@mpe.mpg.de}}

\author{Dieter Lutz\\
        Max-Planck-Institut f\"ur extraterrestrische Physik, Germany\\
        E-mail: \email{lutz@mpe.mpg.de}}

\author{The PEP Consortium\\
http://www.mpe.mpg.de/ir/Research/PEP/index.php}

\abstract{In the era of deep, large-area far-infrared (FIR) surveys from the Herschel Space Telescope, the bulk of 
the star-formation in distant galaxies, once hidden by dust, is now being revealed. The FIR provides probably the cleanest
view of SF in the host galaxies of Active Galactic Nuclei (AGNs) over cosmic time. We report results from studies of the relationships between
SF, AGN activity and AGN obscuration out to $z=2.5$, which employ some of the deepest Herschel and X-ray datasets currently available, 
while spanning orders of magnitude in the dynamic range of AGN properties. We highlight the role of gaseous supply in modulating both
SF and AGN activity without necessarily implying a direct causal connection between these phenomenon. The role of
starburst- or major merger-fueled AGN activity at low and high redshifts is discussed in the context of our results.}

\FullConference{Nuclei of Seyfert galaxies and QSOs - Central engine \& conditions of star formation\\
		 November 6-8, 2012\\
		 Max-Planck-Insitut f\"ur Radioastronomie (MPIfR), Bonn, Germany}

\begin{document}

An understanding of the interplay between the growth of galaxies and the growth of their central supermassive black holes (SMBHs)
is now a fundamental element of galaxy evolution studies. This is due to the remarkable capability of SMBHs to alter
the nature of the gaseous component of their host galaxies through the feedback of accretion
power, and thereby regulate the formation of stars. AGN feedback is now considered a crucial ingredient in modern models of galaxy evolution
\cite{bower06, croton06, sijacki07, somerville08, booth09}, in particular as a pathway for the quenching of star-formation in massive galaxies
and the maintenance of hot gaseous halos at temperatures $>10^{7}$ K for timescales approaching the Hubble time \cite{mcnamara12}.

Most of the local mass density in SMBHs was accreted near the peak of the luminous QSO era at $z\sim2$. Therefore, a
proper treatment of the origin and evolution of local SMBH scaling relationships \cite{magorrian98, tremaine02, graham11} 
needs an understanding of the relationships between star-formation (SF) and AGN activity at that critical epoch, at which
the star-formation rate (SFR) density of the Universe also peaked and when most of the current stellar mass was produced from gaseous raw material \cite{boyle98}. Such relationships are predicted by most models of `causal' galaxy-SMBH co-evolution, i.e 
models that require a direct physical link between the processes that govern SF and those that govern AGN activity. In particular,
an element of synchronisation is key to such models: the SMBH has to grow substantially during a phase in which the galaxy also
builds its stellar mass, or maintain a fixed temporal relationship to this phase. Any lack of synchronisation will smear out 
any causal co-evolutionary processes and undermine the development of tight scaling relationships. 
This may be contrasted to developing concepts that suggest that the simple hierarchical assembly of SMBHs and galaxies 
is sufficient for determining SMBH scaling laws \cite{jahnke11}. In such models, causal co-evolution is not required.

Towards a more constrained observational view, we have undertaken one of the 
largest current studies of the relationship between SF and nuclear activity, spanning local redshifts to $z=2.5$ over
4 orders of magnitude in AGN luminosity. The study builds on the best current X-ray surveys for the identification of AGN
in well-studied extragalactic multi-wavelength fields. SF is probed through the use of far-infrared (FIR) photometry
from the Herschel Space Observatory, building on its unprecedented collecting area -- the largest astronomical mirror in space -- 
and cryogenically-cooled instrumentation in the FIR. For the work reported
here, we primarily rely on data from the Photodetector Array Camera and Spectrometer (PACS) instrument on Herschel,
taken as part of the PACS Evolutionary Probe (PEP) GTO \cite{lutz11} and the GOODS-Herschel GO \cite{elbaz11} programs.  
Our study employs 100 and 160 \mics\ maps in the two GOODS fields (North and South) and the COSMOS field, 
with additional 70 \mics\ data in GOODS-S. Using a combination of prior-based source extraction and stacking of non-detections, 
we derive mean fluxes of various subsamples of sources, typically binned in redshift and at least one other 
quantity such as X-ray luminosity (corrected for absorption in all cases) and X-ray obscuration. From these 
measurements, we interpolate rest-frame mean 60 \mics\ luminosities (\lfir). Several studies, including our own, 
have shown that wavelengths this far in the IR are very weakly affected by AGN-heated 
dust emission, even the stacks of IR-faint samples \cite{netzer07, mullaney11, rosario12}. Measurements in these bands 
track the total IR luminosity and the galaxy-integrated dust-obscured SFR. 

\begin{figure}
\includegraphics[width=0.35\textwidth,angle=90.0]{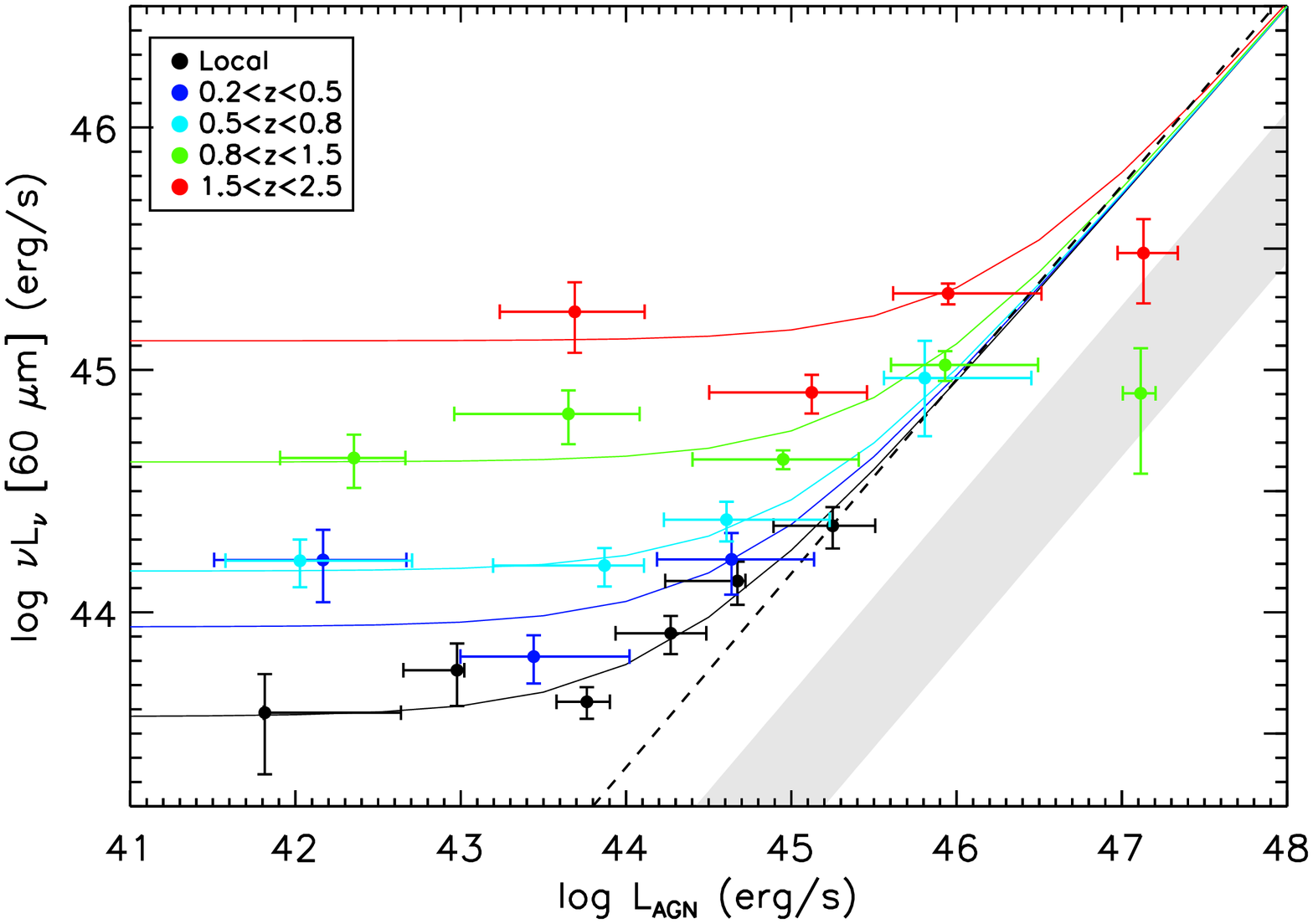}
\includegraphics[width=0.35\textwidth,angle=90.0]{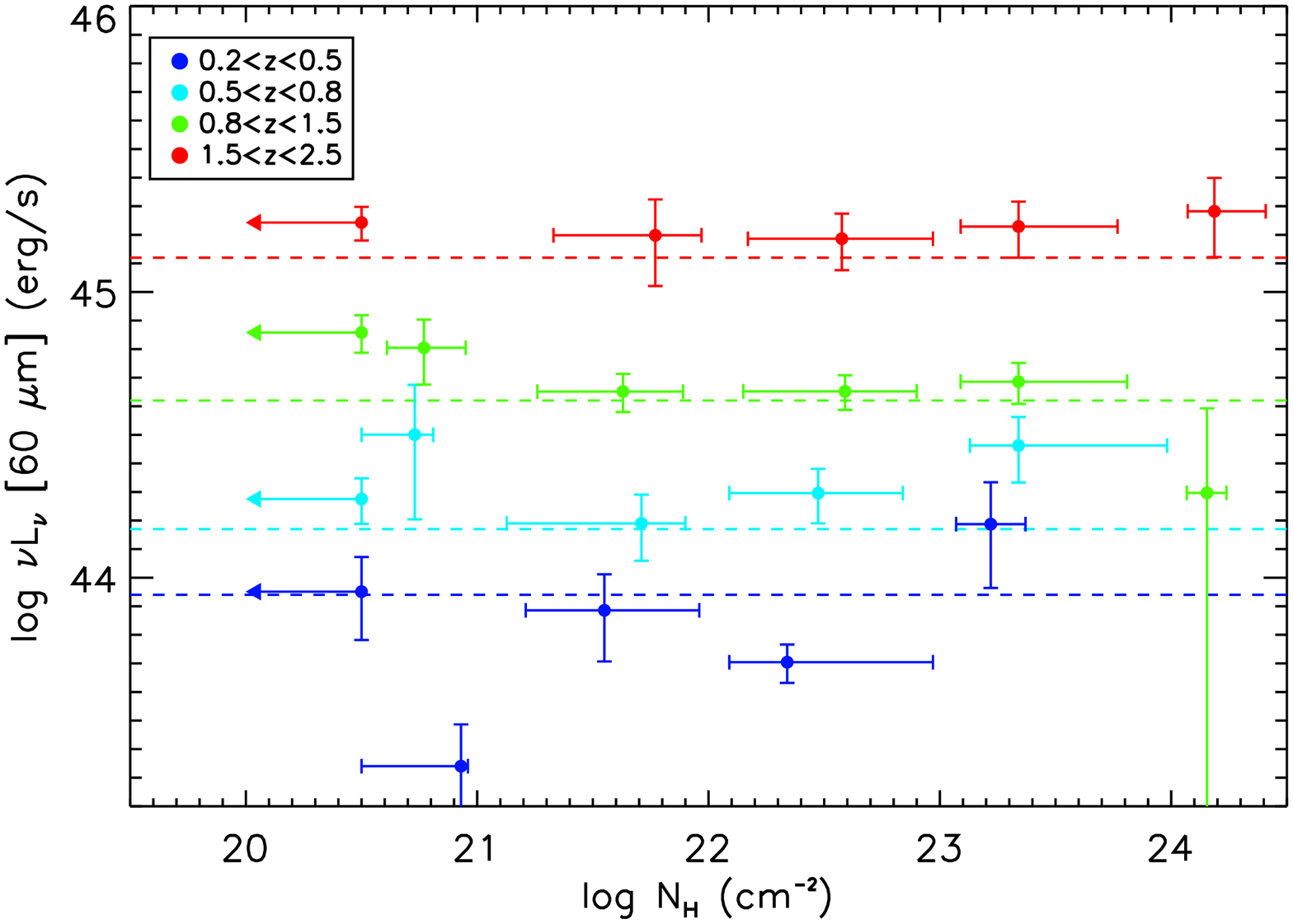}
\caption{{\bf Left:} Trends between the mean rest-frame 60 \mics\ luminosity (\lfir) and 
the bolometric luminosity of the AGN (\lagn). Standard bolometric corrections are used to estimate
\lagn\ from the 2-10 keV X-ray luminosity. \lfir\ is measured in given bins in redshift and \lagn\
by stacking subsamples into PACS maps. In addition to the Herschel
photometry, IRAS-based \lfir\ estimates of the local sample of Swift/BAT AGNs are also included, from the compilation of
\cite{lutz10} and \cite{shao10}. The dashed line is from \cite{netzer09} and the shaded region shows the part
of the diagram occupied by pure AGN SEDs. 
{\bf Right:} Trends between the \lfir\ and the nuclear obscuration of AGNs, expressed as an equivalent Hydrogen 
obscuring column ($N_H$). At all redshifts, only weak correlations are observed, driven more by covariances between 
\lagn\ and $N_H$ in X-ray datasets. See \cite{rosario12} for more details.
}
\label{l60_vs_lagn}
\end{figure}

\section{The Relationship between Star-Formation and AGN power}

Figure \ref{l60_vs_lagn} presents trends between \lfir, a quantity that is roughly proportional to SFR, and
the bolometric luminosity of AGNs  (\lagn). At low AGN luminosities, the mean SFR of X-ray selected
AGNs is independent of \lagn\ at all redshifts, but rises steadily
with redshift in a manner that tracks the mean SFR of massive inactive galaxies \cite{wuyts11}. At $z<1$, 
following a characteristic turnover at \lagn$> 10^{44-45}$ \ergs, a correlation is seen between \lfir\ and AGN
output at high nuclear luminosities. We interpret this turnover as the increased importance of starburst or merger-driven
AGN fueling at these luminosities, since models of these mechanisms predict such a correlation \cite{rosario12}.  
At higher redshifts, the characteristic correlation weakens or disappears and a flat trend is seen across all AGN 
luminosities. While still to be securely confirmed, the lack of a strong trend suggests that, at higher redshifts, 
disk-instability mediated fueling of even rather luminous AGNs (i.e, fast growing black holes, \cite{bournaud11})
may be more important at the expense of merger-driven fueling or other mechanisms that synchronize SF and AGN activity.

\begin{figure}
\includegraphics[width=1.0\textwidth]{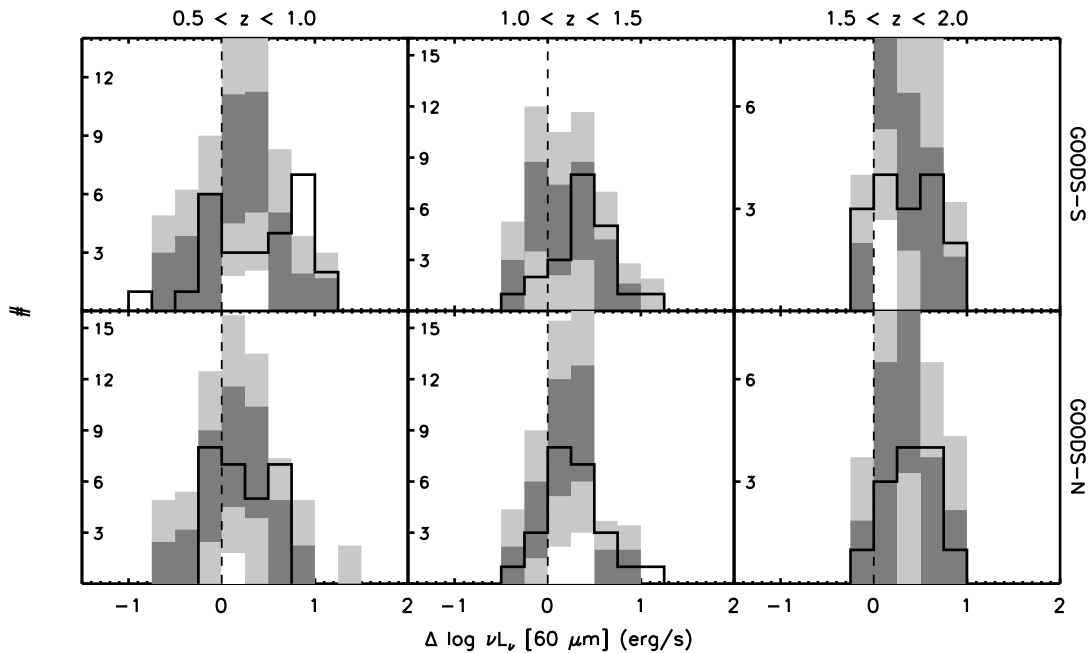}
\caption{A comparison of the distribution of SFR offsets from the Main Sequence between AGNs (open histograms) and mass-matched
inactive control galaxies (shaded histograms) in the two GOODS fields. The shadings in the latter histograms represent the
scatter in the SFR distributions from 1000 bootstrap samples drawn from the inactive galaxy population: dark/light histograms
are the $1/2\sigma$ scatter. The dashed line is the zero offset: the ridge line of the Main Sequence. The AGNs show
SFR distributions that are statistically equivalent to inactive galaxies. See \cite{rosario13b} for more details.}
\label{lfir_dists}
\end{figure}

\begin{figure}
\includegraphics[width=1.0\textwidth]{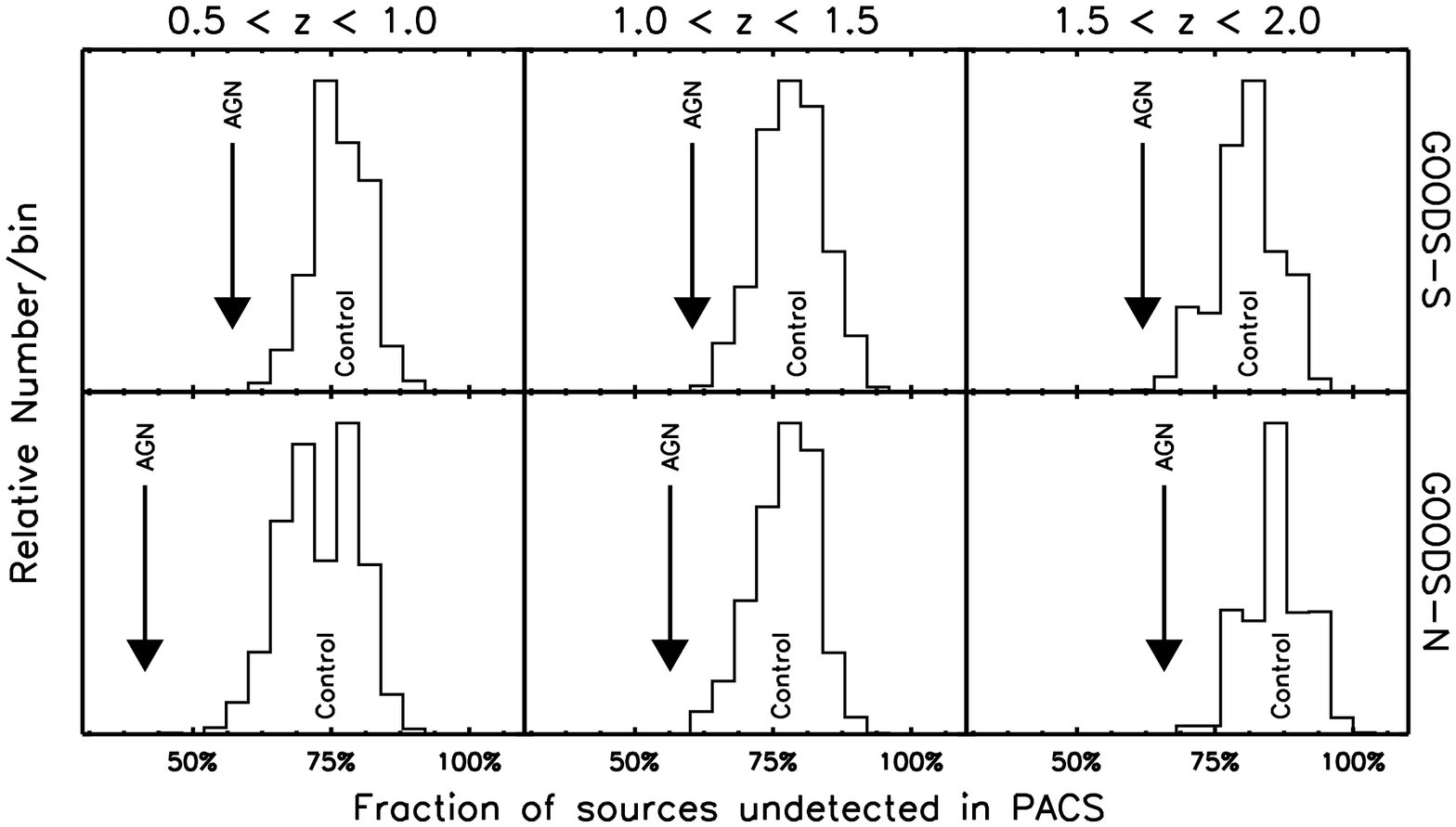}
\caption{AGNs are more likely to be SF hosts than massive galaxies of the same stellar mass. Here we compare, for AGNs and
mass-matched inactive control galaxies, the fractions of sources \emph{not detected} in deep Herschel/PACS data. This fraction
encompasses all quenching and quiescent galaxies in these samples. The histogram for the control sample comes
from 1000 bootstrap samples drawn from the inactive galaxy population, which may be compared to the arrow which 
shows the fraction among X-ray AGNs. Typically 75-80\%\ of massive galaxies lie below the PACS detection limit, while
only 55-60\%\ of AGN do so.}
\label{detfracs}
\end{figure}

\section{The Supply of Cold Gas modulates Low and Moderate AGN Activity}

SF galaxies display a relatively tight relationship between SFR and stellar mass (\smass), popularly known
as the SF Main Sequence \cite{noeske07, wuyts11, whitaker12}. After accounting for the effects of
AGN contamination on host stellar mass estimates, \cite{santini12} find that \lfir\ in AGNs also tracks
\smass\ in a similar fashion as inactive galaxies. However, a small enhancement of $\approx 0.2$ dex
is found in the mean SFR of AGNs over inactive galaxies, which is at apparent odds with the
very similar distributions of SFRs found from FIR studies of star-forming galaxies and AGNs \cite{mullaney12a}. In \cite{rosario13b}, 
we studied the SFR distributions and PACS detection rates of AGNs and inactive galaxies, taking
into account the unusual mass function of AGN hosts, which differs from the general inactive galaxy sample
\cite{kauffmann03, silverman09, rosario13a}. This was done using the deepest current FIR data available
in any extragalactic field: the PEP+GOODS-Herschel combined maps. The SFR distributions of low-to-moderate luminosity
AGNs are statistically indistinguishable from inactive galaxies in both fields and at all redshifts (Figure \ref{lfir_dists}), 
confirming the results of \cite{mullaney12a}. However, the PACS detection rates of AGN hosts are always higher 
than equally massive inactive galaxies at $> 3\sigma$ significance, implying that AGNs are most likely
to be in SF hosts than in passive hosts. In particular, we show that the high incidence of AGN in the `Green Valley'
is not a consequence of a link between AGN activity and the quenching of their hosts, but mostly driven by mass
selection effects. AGN are in fact found less frequently in quenching or quiescent galaxies. We attribute this to the low 
supply of cold gas in such systems, since cold gas are required for both the fueling of radiatively efficient modes of AGN activity
as well as current SF.

\section{The Relationship between Star-Formation and AGN obscuration}

A key prediction of some models of co-evolution is a close association between starburst activity in the host
galaxy and obscured black hole growth. A search for trends between \lfir\ and the X-ray obscuration towards
the nucleus finds essentially no relationships between them out to $z\sim2.5$ (Figure \ref{l60_vs_lagn}). Weak correlations 
visible in such a diagram are primarily driven by selection effects, since only the most luminous obscured AGNs are detectable in X-ray
surveys.  

\section{Implications for the Co-evolution of Galaxies and Black Holes}

Causal or synchronized models of the co-evolutionary connection of SMBHs and their host galaxies predict strong correlations
between SF and AGN activity. Our studies show that these correlations are weak among low and moderate luminosity AGNs, suggesting
that strong co-evolution does not occur in these systems. More importantly, we provide some evidence for the notion that synchronized
growth is weak at high redshifts even among luminous AGNs. However, our study of low and moderate luminosity AGNs suggests an
important, though indirect, channel for co-evolution. As a galaxy accretes gas from the cosmic web, some fraction of this gas
stochastically and intermittently 
falls into the SMBH through normal secular inflow processes. SMBHs grow faster in galaxies with more cold gas, but it is
precisely these galaxies that also have a higher mean SFR. SMBH scaling relations, and correlations between mean SMBH
accretion rate and stellar mass \cite{mullaney12b}, may be mediated more by the availability or supply of fuel rather
than causal links between AGN fueling and star-formation.

\end{document}